\documentclass[preprint]{aastex}

\usepackage{graphicx}
\usepackage{array}
\usepackage{natbib}
\usepackage{amssymb}
\usepackage{lscape}

\begin{document}

\title{\textit{Spitzer} Point-Source Catalogs of $\sim300,000$ Stars in Seven Nearby Galaxies\altaffilmark{1}}

\author{Rubab~Khan\altaffilmark{2,3},
K.~Z.~Stanek\altaffilmark{4,5},
C.~S.~Kochanek\altaffilmark{4,5},
G.~Sonneborn\altaffilmark{3}
}

\altaffiltext{1}{Based on observations made with the {\it Spitzer} Space Telescope, 
which is operated by the Jet Propulsion Laboratory, California Institute of Technology 
under a contract with NASA.}

\altaffiltext{2}{JWST Fellow, NASA Postdoctoral Program, ORAU,
P.O. Box 117, MS 36, Oak Ridge, TN 37831}

\altaffiltext{3}{NASA Goddard Space Flight Center, MC 665,
8800 Greenbelt Road, Greenbelt, MD 20771; rubab.m.khan, george.sonneborn-1@nasa.gov}

\altaffiltext{4}{Dept.\ of Astronomy, The Ohio State University, 140
W.\ 18th Ave., Columbus, OH 43210; kstanek, ckochanek@astronomy.ohio-state.edu}

\altaffiltext{5}{Center for Cosmology and AstroParticle Physics, 
The Ohio State University, 191 W.\ Woodruff Ave., Columbus, OH 43210}

\shorttitle{\textit{Spitzer} Catalog of $7$ Galaxies}

\shortauthors{Khan et al. 2015(b)}

\begin{abstract}
\label{sec:abstract}
We present \textit{Spitzer} IRAC $3.6-8\,\micron$ and MIPS $24\,\micron$ point-source 
catalogs for seven galaxies: NGC\,$6822$, M\,$33$, NGC\,$300$, NGC\,$2403$, M\,$81$, 
NGC\,$0247$, and NGC\,$7793$. The catalogs contain a total of $\sim300,000$ sources and were 
created by dual-band selection of sources
with $>3\sigma$ detections at both $3.6\,\micron$ and $4.5\,\micron$.
The source lists become significantly incomplete near $m_{3.6}=m_{4.5}\simeq18$.
We complement the $3.6\,\micron$ and $4.5\,\micron$ fluxes with $5.8\,\micron$, $8.0\,\micron$ and 
$24\,\micron$ fluxes or $3\sigma$ upper limits using a combination of PSF and 
aperture photometry. This catalog is a resource as an archive for studying mid-infrared transients
and for planning observations with the James Webb Space Telescope.
\end{abstract} 

\maketitle

\section{Introduction}
\label{sec:intro}

The {\it Spitzer} Space Telescope \citep[{\it Spitzer},][]{ref:Werner_2004}
enabled mid-infrared (mid-IR) observations with unprecedented sensitivity. 
An enormous archive of imaging data has been collected using the 
Infrared Array Camera~\citep[IRAC,][]{ref:Fazio_2004} and the Multiband 
Imaging Photometer~\citep[MIPS,][]{ref:Rieke_2004} instruments aboard 
{\it Spitzer} that have been utilized to study astrophysical objects in 
the Galaxy \citep[e.g.,][]{ref:Benjamin_2003} and beyond.
In particular, {\it Spitzer} made it possible to 
study resolved stellar 
populations in the Magellanic Clouds~\citep{ref:Meixner_2006,ref:Bolatto_2007,ref:Gordon_2011}
and local group galaxies such as M\,31~\citep{ref:Barmby_2006,ref:Mould_2008} and M\,33
\citep{ref:McQuinn_2007,ref:Thompson_2009} in the mid-IR. 

Although global galaxy properties
beyond the local group have been extensively studied 
using {\it Spitzer} images \citep{ref:Kennicutt_2003,ref:Dale_2009,ref:Sheth_2008}, efforts to catalog individual mid-IR luminous sources in 
these galaxies have been limited. There are a number of sources of archival {\em Spitzer} data for nearby 
galaxies. The \textit{Spitzer} Infrared Nearby Galaxies Survey~\citep[SINGS,][]{ref:Kennicutt_2003} 
made a comprehensive mid-IR imaging and 
spectroscopic survey of 75 galaxies, many of them within $10\;$Mpc. The 
Local Volume Legacy Survey~\citep[LVL,][]{ref:Dale_2009} surveyed a total of 256 nearby galaxies, including 
all known galaxies inside a sub-volume bounded by $3.5\;$Mpc and an unbiased 
sample of S-Irr galaxies within a larger, and more representative, $11\;$Mpc 
sphere. The {\it Spitzer} Survey of Stellar Structure in Galaxies 
\citep[$S^4G$,][]{ref:Sheth_2008} collected data for $\sim2300$ 
galaxies within $40\;$Mpc using the warm \textit{Spitzer} ($3.6\,\micron$ and $4.5\,\micron$) bands. 

While it is difficult to identify and characterize mid-IR point sources in the crowded 
and dusty disks of large star forming galaxies due to IR emission from interstellar 
dust, blending and background contamination, it is nonetheless feasible to identify 
and photometer mid-IR luminous stars in galaxies well beyond the Magellanic clouds
\citep{ref:Thompson_2009,ref:Khan_2010,ref:Khan_2011,ref:Gerke_2012}.
In \citet{ref:Khan_2013},
we used archival IRAC images of seven galaxies 
($\lesssim4$\,Mpc; closest to farthest: NGC\,$6822$, M\,$33$, NGC\,$300$ 
\citep[see][]{ref:Helou_2004}, NGC\,$2403$, M\,$81$ \citep[see][]{ref:Willner_2004}, NGC\,$0247$, 
and NGC\,$7793$) in a pilot study to search for extragalactic analogs of the Galactic 
object $\eta$\,Carinae, taking advantage of 
the data made available by the SINGS and LVL projects, which led to 
the identification of an emerging class of evolved massive 
($M\simeq25\sim60\,M_\odot$) stars \citep{ref:Khan_2015}.

Here we present photometric inventories of the mid-IR point sources in 
the IRAC $3.6\,\micron$, $4.5\,\micron$, $5.8\,\micron$, $8\,\micron$ and MIPS $24\,\micron$ images of the 
galaxies studied by \citet{ref:Khan_2013,ref:Khan_2015}. 
Although we concentrated on galaxies with recent star formation, as 
only these would have large numbers of the short lived, very massive 
stars that were our primary targets, we also included 
the small, low-mass galaxy NGC\,6822 
as a test for examining large numbers of smaller, lower-metallicity systems. 
M\,33~\citep[D\,$\simeq0.96$\,Mpc,][]{ref:Bonanos_2006} was previously 
cataloged by \citet{ref:McQuinn_2007} in the IRAC $3.6$, $4.5$ and $8.0\,\micron$ bands, and 
by \citet{ref:Thompson_2009} in the IRAC $3.6$ and $4.5\,\micron$ bands.
Point sources in NGC\,300 \citep[D\,$\simeq1.9$\,Mpc,][]{ref:Gieren_2005} and 
M\,81 \citep[D\,$\simeq3.6$\,Mpc,][]{ref:Gerke_2011} 
were cataloged by \citet{ref:Khan_2010} in the IRAC $3.6$ and $4.5\,\micron$ bands.
The catalogs presented in this paper identify a larger number of sources in these galaxies 
than the previous studies.
{\it Spitzer} point-source catalogs of NGC\,6822 \citep[D\,$\simeq0.46$\,Mpc,][]{ref:Gieren_2006}, 
NGC\,2403 \citep[D\,$\simeq3.1$\,Mpc,][]{ref:Saha_2006}, 
NGC\,0247 \citep[D\,$\simeq3.6$\,Mpc,][]{ref:Madore_2009} and 
NGC\,7793 \citep[D\,$\simeq4.1$\,Mpc,][]{ref:Tully_2009} are being 
published here for the first time.

We use the same mosaics that were utilized by \citet{ref:Khan_2013,ref:Khan_2015}.
For M\,33, we use the six co-added epochs of IRAC images from~\citet{ref:McQuinn_2007} that 
was originally created and used by \citet{ref:Thompson_2009}, and 
the MIPS data retrieved from the 
\textit{Spitzer Heritage Archive}\footnote[6]{\tt http://sha.ipac.caltech.edu/applications/Spitzer/SHA/}. 
For NGC\,300 and NGC\,247, we used the mosaics from the LVL survey \citep{ref:Dale_2009}. For NGC\,6822, NGC\,2403, M\,81, 
and NGC\,7793, we used the mosaics from the SINGS survey~\citep{ref:Kennicutt_2003}. We utilize the 
full mosaics available for each galaxy. Figure~\ref{fig:galaxies} shows the IRAC 3.6\,$\mu$m
images of the targeted galaxies.
In what follows, we describe our methodology 
(Section\,\ref{sec:photo}) and present the point-source catalogs 
(Section\,\ref{sec:cats}).

\section{Photometry}
\label{sec:photo}

In this Section, we detail how we obtained the photometric 
measurements at various wavelengths and combined them to construct the 
point-source catalogs. Although the procedures followed here are derived 
from the techniques developed by \citet{ref:Khan_2010} and 
\citet{ref:Khan_2013}, there are some key differences, as we now 
carry out an inventory of all point sources
rather than targeting a particular sub class with 
desired photometric properties. 

Specifically, \citet{ref:Khan_2010} searched 
for very red mid-IR ($m_{3.6}-m_{4.5}>1.5$) sources near or at the detection limit 
of the first two IRAC bands, and therefore included in their primary 
source-list objects that were detected in the $4.5\,\micron$ but
not the $3.6\,\micron$ image, as well as objects that could only be detected in 
the $4.5\micron - 3.6\micron$ difference image but not in either of the individual images. 
\citet{ref:Khan_2013} focused on 
the most luminous mid-IR ($L_{mIR}>10^{5}\,L_\odot$) sources with a 
spectral energy distribution (SED) that is either flat across the four IRAC 
bands or rising towards the longer wavelengths, and therefore included in 
the primary source list all objects 
that had $\lambda L_\lambda>10^{4}\,L_\odot$ in any of the first three 
($3.6\,\micron, 4.5\,\micron, 5.8\,\micron$) IRAC bands. 

In this work, we implement strict detection criteria by 
selecting all sources detected at $>3\sigma$
in both the $3.6\,\micron$ and $4.5\,\micron$ images within a certain 
matching radius as point sources. Next, 
we search for $>3\sigma$ detections of these point sources 
in the $5.8\,\micron$ and $8.0\,\micron$ images within the same matching 
radius. If no counterpart is found, we attempt 
to measure the flux at the location of the $3.6/4.5 \micron$ point source through PSF fitting, 
and failing that, through aperture photometry. For the MIPS 
$24\,\micron$ images, we only use aperture photometry 
due to the much lower resolution and larger PSF size compared to the IRAC images. 
Finally, for all objects that do not have a $>3\,\sigma$ detection 
at $5.8\,\micron$, $8.0\,\micron$ and $24\,\micron$, we estimate the $3\sigma$ flux upper
limits. The fluxes and upper limits are transformed to Vega-calibrated magnitudes using 
the flux 
zero points\footnote[7]{$280.9$, $179.7$, $115.0$, $64.13$ and $7.17$\,Jy for $3.6$, $4.5$, $5.8$, $8.0$ and $24\,\micron$ bands.}
and aperture corrections provided in the \textit{Spitzer} Data Analysis 
Cookbook\footnote[8]{\tt http://irsa.ipac.caltech.edu/data/SPITZER/docs/dataanalysistools/}. 
Given this broad outline, we now describe the specific technical details of how we 
performed the measurements at the various stages of constructing the catalogs.

\begin{figure}
\begin{center}
\includegraphics[angle=0,width=120mm]{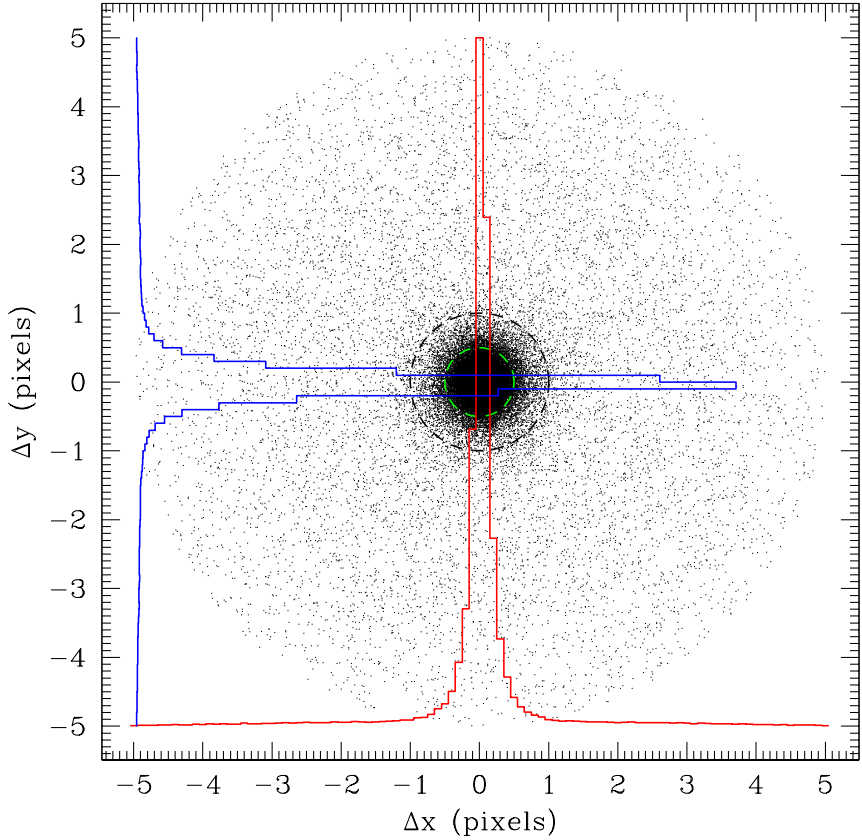}
\end{center}
\caption{The distances (in pixels) to the nearest $3.6\,\micron$ 
source for each $4.5\,\micron$ source in M\,33 along the x axis 
($\Delta x$) and y axis ($\Delta y$) of the images. The blue and 
red histograms show the distance distributions. The radius of the 
concentric circles are 0.5\,pixel (green) and 1\,pixel (black). In this case,
most sources  ($>90\%$) 
have a match within $0.5$\,pixels, and the density of 
nearest matches falls rapidly between $0.5-1.0$\,pixel ($<10\%$ additional 
matches) while the number of duplicate matches increase 
($<0.5\%$ duplicate matches). At larger distances, the distribution essentially flattens. 
Similiar distributions are observed for the other six galaxies 
(see Table\,\ref{tab:stats}).}
\label{fig:match}
\end{figure}

We used the DAOPHOT/ALLSTAR PSF-fitting and photometry package~\citep{ref:Stetson_1992} 
to construct the PSFs, to identify the $>3\sigma$ sources, and to measure their 
flux at all $4$ IRAC bands. The different roll angles of the various {\it Spitzer} observations
made it necessary to construct
the PSFs for each galaxy in each band independently.
Next, we empirically determined the optimal radius to match the $3.6\,\micron$ and 
$4.5\,\micron$ source lists. Figure\,\ref{fig:match} shows the distribution of 
distances to the nearest $3.6\,\micron$ source for each $4.5\,\micron$ source in 
M\,33. In this case, over $90\%$ 
have a match within $0.5$\,pixel. The density of 
nearest matches falls rapidly between $0.5-1.0$\,pixel ($<10\%$ additional 
matches), while the number of duplicates increases 
($<0.5\%$ duplicate matches), and then the distribution essentially flattens. 
Similar distributions are observed for the other six galaxies  
(see Table\,\ref{tab:stats}). We therefore adopted an empirically 
motivated matching radius of 1\,pixel in order to maximize 
the number of matches for a minimal number of chance superpositions.

We used the IRAF\footnote[9]{IRAF is distributed by the 
National Optical Astronomy Observatory, which is operated by the Association of 
Universities for Research in Astronomy (AURA) under cooperative agreement with 
the National Science Foundation.} ApPhot/Phot tool for performing aperture photometry
at the point-source locations for all IRAC bands and the MIPS $24\,\micron$ band.
For the four IRAC bands, we use an extraction aperture of $2\farcs4$, a local background 
annulus of $2\farcs4 - 7\farcs2$, and aperture corrections of $1.213$, $1.234$, $1.379$, 
and $1.584$ respectively.
For the MIPS $24\,\micron$ band, we use an extraction aperture of $3\farcs5$, a local background 
annulus of $6\farcs - 8\farcs$, and an aperture correction of $2.78$.
We estimate the local background using a $2\sigma$ outlier rejection procedure in 
order to exclude sources located in the local sky annulus, and correct for the 
excluded pixels assuming a Gaussian background distribution. Using a 
background annulus immediately next to the signal aperture minimizes the effects of 
background variations in the crowded fields of the galaxies. We also determine 
the $3\sigma$ flux upper limit for each aperture location using the 
local background estimate.

\begin{figure}
\begin{center}
\includegraphics[angle=0,width=160mm]{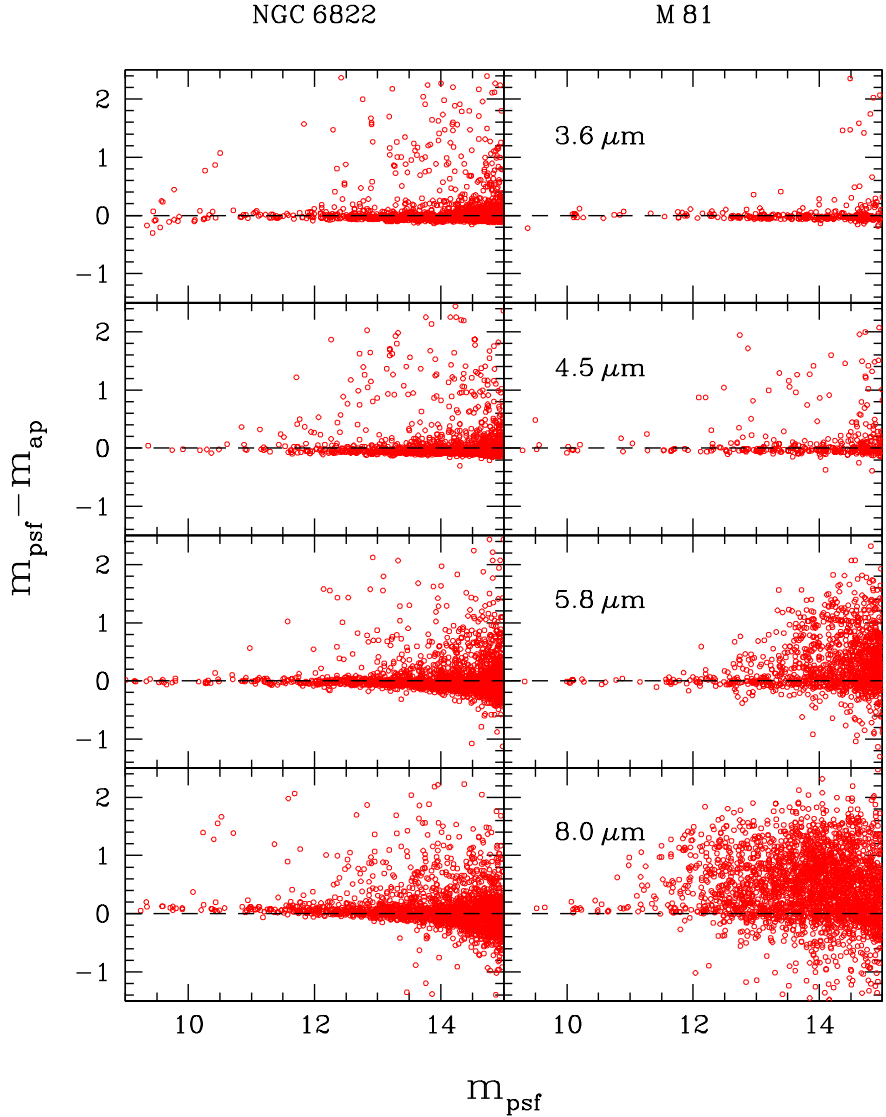}
\end{center}
\caption{The differences between aperture and PSF photometry magnitudes
as a function of the PSF magnitudes for the $m_{psf}<15$
sources in NGC\,6822 and M\,81. For the less 
crowded case of NGC\,6822, the two measurements generally agree 
in all four IRAC bands, with the scatter increasing 
for the fainter sources. The same is true for the $3.6\,\micron$ and 
$4.5\,\micron$ images of M\,81, but at $5.8\,\micron$, the scatter 
increases rapidly, and for 
$8.0\,\micron$ the situation worsens further. 
We found this to be true for the other galaxies and M\,81 
shows behavior that is representative of all the targets apart from NGC\,6822.}
\label{fig:offset}
\end{figure}

Ideally, flux measurements of an isolated point source through either aperture or 
PSF photometry would produce the same 
results after appropriate aperture corrections (in the first case)
and small zero-point offsets (in the second case) to account for flux 
underestimation due to PSF fitting up to finite radius rather than to infinity. 
We derive this small (usually $\sim0.1$\,mag) offset for each image from the mean 
difference between the magnitudes of relatively isolated, unsaturated bright sources measured through aperture and PSF 
photometry. For fainter sources, especially in limited spatial 
resolution images of crowded fields, the aperture and PSF photometry measurements can vary significantly. 
Figure\,\ref{fig:offset} shows the differences between apparent magnitudes 
determined through aperture and PSF photometry as a function of PSF magnitude
for the $m_{psf}<15$ sources in NGC\,6822 and M\,81. 
For the less crowded case of NGC\,6822, the two measurements generally agree for the very 
brightest sources in all four IRAC bands, with the scatter increasing 
for the fainter sources. The same is true for the $3.6\,\micron$ and 
$4.5\,\micron$ images of M\,81, but at $5.8\,\micron$, the scatter 
is much worse, although a sequence of bright sources with good agreement 
between the two sets of measurements can still be seen. However, for 
$8.0\,\micron$, such a sequence cannot be clearly identified.
We found this to be true for 
the other galaxies and M\,81 
shows behavior that is representative of all the targets apart from NGC\,6822. Mismatches between the two magnitudes are a 
good indicator of when crowding is significantly effecting the magnitude estimates, 
and in the catalogs we list the difference 
between the PSF and aperture photometry magnitudes for each source.

Because of these crowding problems, 
we do not attempt to fine tune the $8.0\,\micron$ PSF photometry 
measurements by applying the small 
linear offset, although 
we do this for the other three IRAC bands. For all IRAC bands, we 
universally prefer PSF photometry over aperture photometry, 
because PSF fitting is more successful at extracting relatively fainter 
sources in crowded fields. By definition, the nature of PSF fitting 
is that it takes into account 
whether the flux distribution of a potential point source 
is consistent with the PSF shape, and the degree of agreement or disagreement 
is reflected by a smaller or larger photometric uncertainty. Rather than conducting a 
bulk accounting of all excess flux within an aperture, PSF fitting measures a 
weighted sum of excess flux within a 
finite aperture. This is significant for fainter sources in crowded 
fields that are often close to brighter sources. PSF fitting more accurately 
identifies the flux associated with such sources isolating the contamination from 
other sources as well as more accurately subtracting the local sky.
On the other hand, aperture photometry can 
significantly over subtract the sky and underestimate the source flux, 
or overestimate the source flux by failing to remove 
contamination from nearby brighter sources. Nevertheless, aperture photometry proves very useful 
for validating the PSF photometry measurements at the bright end in all IRAC bands, 
for estimating $5.8\,\micron$ and $8.0\,\micron$ fluxes 
where PSF fitting fails and $24\,\micron$ flux
where the lower resolution makes PSF photometry infeasible,
and for determining flux upper limits.

To summarize, we implement strict detection criteria by requiring a $>3\sigma$ 
detection of all cataloged sources at $3.6\,\micron$ and $4.5\,\micron$. 
We then complement those measurements 
at the $5.8\,\micron$, $8.0\,\micron$ and $24\,\micron$ bands through a combination of PSF and 
aperture photometry, preferring PSF fitting over aperture photometry at  
$5.8\,\micron$ and $8.0\,\micron$, and  exclusively using aperture photometry at  
$24\,\micron$. For all objects that do not have a $>3\,\sigma$ detection 
at these three longer wavelengths, we estimate $3\sigma$ flux upper
limits.

\section{Catalogs}
\label{sec:cats}

In this Section, we discuss the results of our mid-IR photometric survey.
Because we required a $>3\sigma$ 
detection for each source at $3.6\,\micron$ and $4.5\,\micron$, the effective survey area for each 
galaxy is the overlap of the IRAC $3.6\,\micron$ and $4.5\,\micron$
image mosaics.
Table\,\ref{tab:stats} lists the effective 
survey area, gas phase (H$\alpha$) star formation rates adopted from \citet{ref:Khan_2013} and
the number of point sources cataloged in each galaxy followed by the 
number of matches at $<0.5$ and $0.5-1.0$ 
pixel, duplicates 
between the $3.6\,\micron$ and $4.5\,\micron$
source lists, and the number of $>3\sigma$ counterparts identified 
at the three longer wavelength images for each galaxy.
Tables\,$2-8$ list the coordinates (J$2000.0$; RA and Dec) of the 
point sources followed by their Vega calibrated apparent magnitudes ($m_\lambda$), the associated 
$1\,\sigma$ uncertainties ($\sigma_\lambda$), and (for the $3.6-8.0\,\micron$ bands) the differences 
between the PSF and aperture photometry magnitudes ($\delta_\lambda$).
For the $5.8\,\micron$, $8.0\,\micron$ and $24\,\micron$ bands, 
$\sigma_\lambda=99.99$ implies that the associated photometric measurement is a
$3\,\sigma$ flux upper limit, and $m_\lambda=99.99$ (as well as 
$\sigma_\lambda=99.99$) indicates that no reliable photometric measurement could be 
obtained for that location. For the IRAC bands, $\delta_\lambda=99.99$ implies 
that one or both of the associated photometric
measurements did not yield a $>3\sigma$ flux measurement.
 
\begin{figure}
\begin{center}
\includegraphics[angle=0,width=160mm]{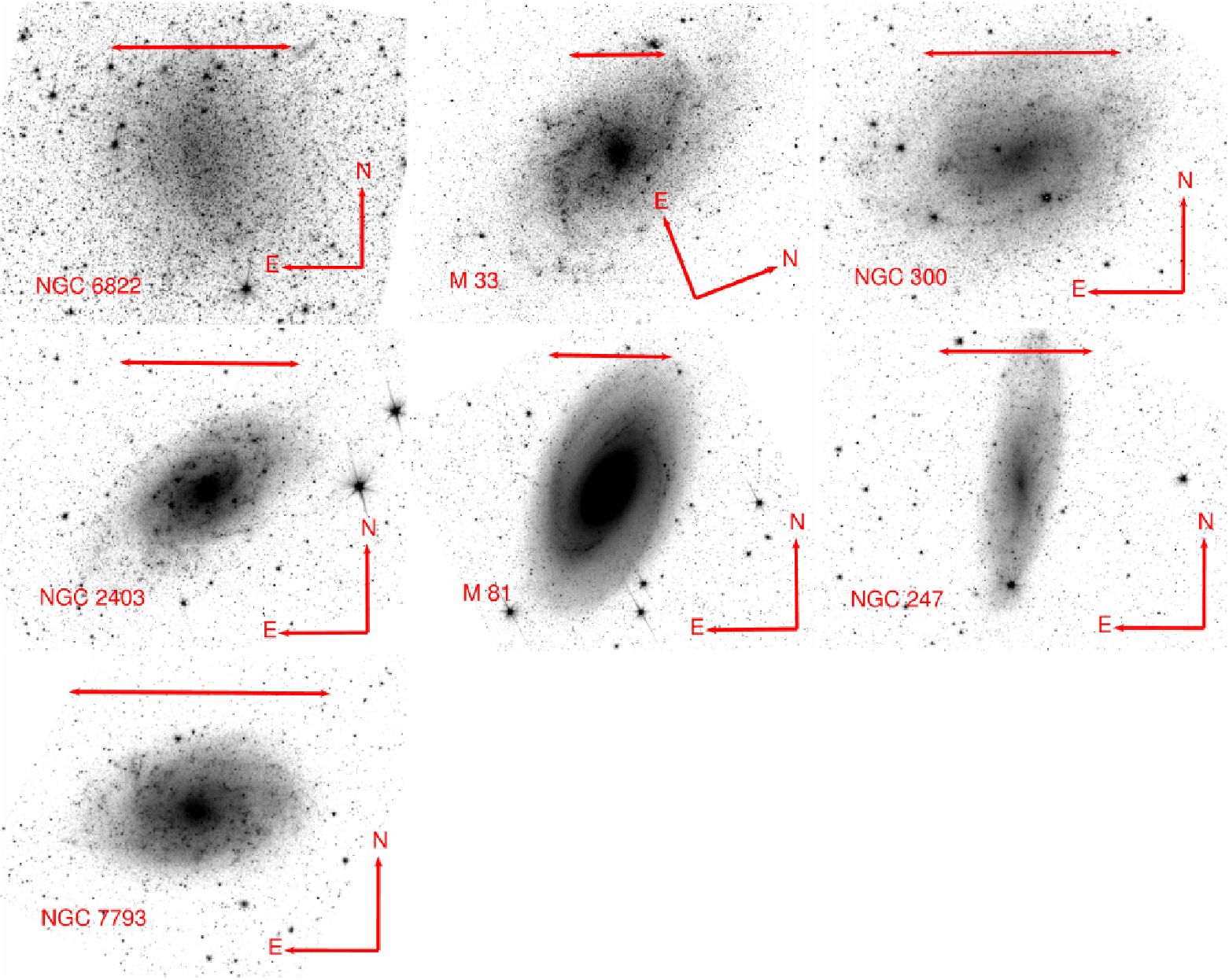}
\end{center}
\caption{The IRAC 3.6\,$\mu$m images of the seven targeted galaxies: 
NGC\,6822, M\,33, NGC\,300, NGC\,2403, M\,81, NGC\,0247, 
and NGC\,7793. The red line on each figure spans $10\farcm$}
\label{fig:galaxies}
\end{figure}

\begin{figure}
\begin{center}
\includegraphics[angle=0,width=160mm]{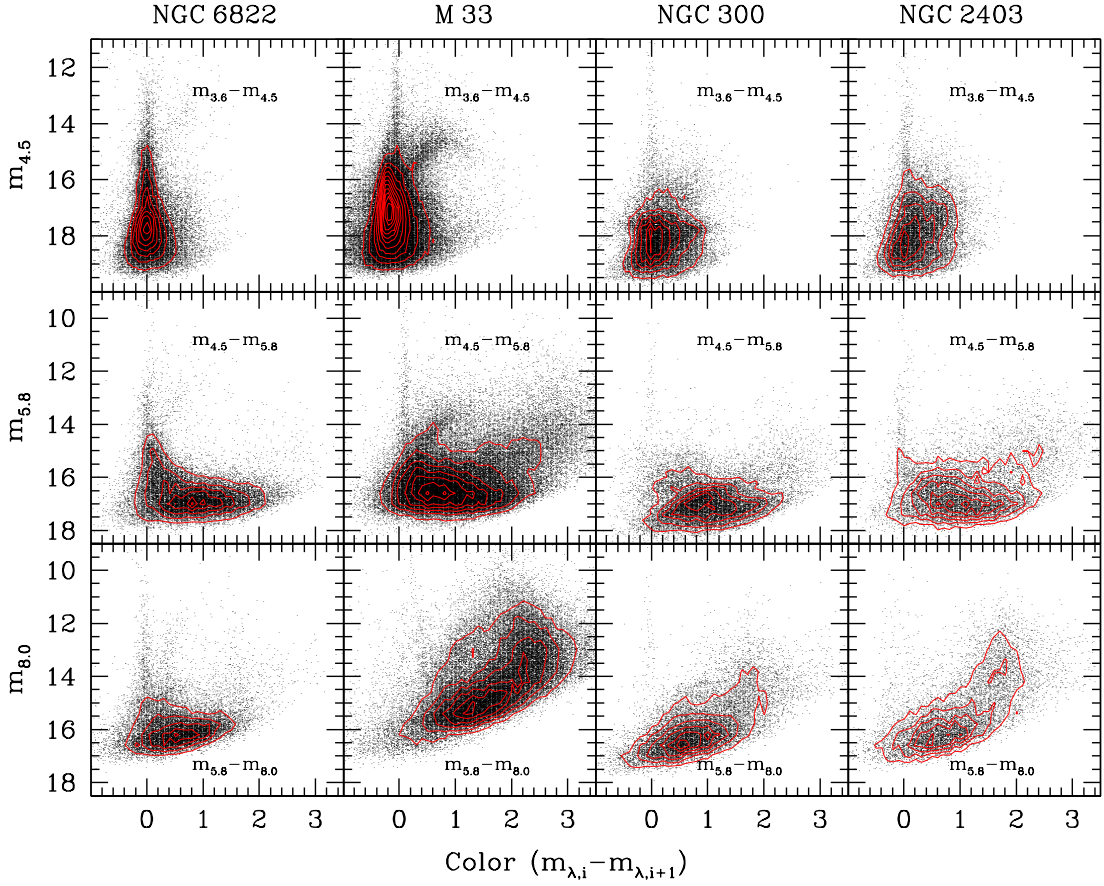}
\end{center}
\caption{The $m_{4.5}$\,vs.\,$m_{3.6}-m_{4.5}$ (top row), $m_{5.8}$\,vs.\,$m_{4.5}-m_{5.8}$ (middle row), and 
$m_{8.0}$\,vs.\,$m_{5.8}-m_{8.0}$ (bottom row) color magnitude diagrams (CMDs) for the 
$>3\sigma$ sources in NGC\,6822, M\,33, NGC\,300 and NGC\,2403. The red lines 
represent isodensity contours.}
\label{fig:cmd16a}
\end{figure}

\begin{figure}
\begin{center}
\includegraphics[angle=0,width=160mm]{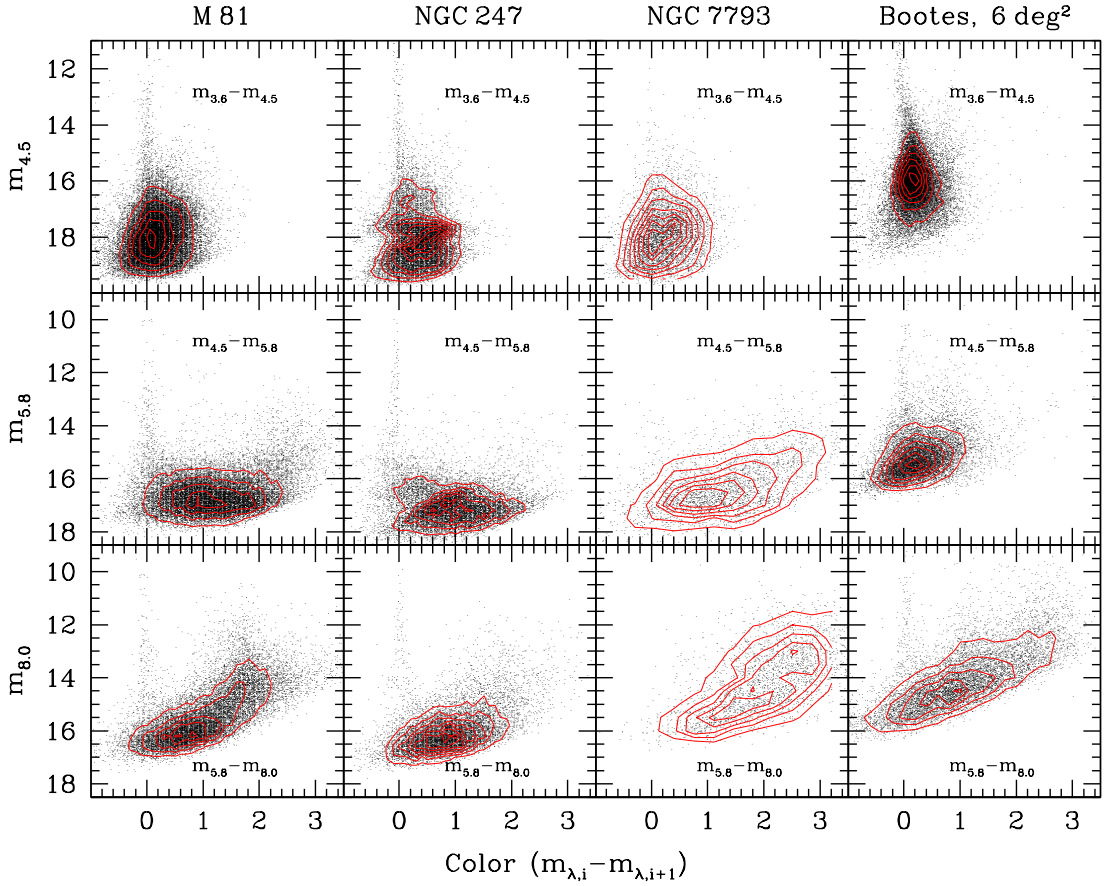}
\end{center}
\caption{Same as Figure\,\ref{fig:cmd16a}, but for the galaxies M\,81,
NGC\,247, NGC\,7793, and (for comparison) a $6$\,deg$^2$ region 
of the NOAO Bootes Field  
from the {\it Spitzer} Deep Wide Field Survey 
\citep[SDWFS,][]{ref:Ashby_2009} data.}
\label{fig:cmd16b}
\end{figure}

Figures \ref{fig:cmd16a} and \ref{fig:cmd16b} present the 
$m_{4.5}$\,vs.\,$m_{3.6}-m_{4.5}$, $m_{5.8}$\,vs.\,$m_{4.5}-m_{5.8}$, and 
$m_{8.0}$\,vs.\,$m_{5.8}-m_{8.0}$ color magnitude diagrams (CMDs) for 
each galaxy. For comparison, we include the mid-IR CMDs for all 
sources in a $6$\,deg$^2$ region \citep[see][for details]{ref:Khan_2013} 
of the NOAO Bootes Field produced 
from the {\it Spitzer} Deep Wide Field Survey 
\citep[SDWFS,][]{ref:Ashby_2009} data. Our catalogs
simply inventory all the sources present on the image mosaics
and do not attempt to 
distinguish between sources actually associated with the 
galaxies and unrelated contaminants. The contamination
is significant for galaxies like NGC\,247, which is highly inclined
and covers a smaller fraction of the {\it Spitzer} images,
than for the larger and more face-on galaxies (see 
Figure\,\ref{fig:galaxies}). Indeed, the CMDs of NGC\,247 
(Figure\,\ref{fig:cmd16b}, second column) clearly show two 
distinct population of sources, with the break appearing 
near $m_{4.5}\simeq18$. The CMDs of the two closest galaxies 
(NGC\,6822 and M\,33) show distinguishable sequences of 
bright and red AGB stars (near $m_{4.5}\simeq15$), but for the more 
distant galaxies this feature is less prominent (see \citealp{ref:Khan_2010} 
for relevant discussion). The CMDs of NGC\,7793
has the fewest identified point sources both due to its lower mass 
and because of poorer quality (more systematic artifacts) of its image
mosaics.

\begin{figure}
\begin{center}
\includegraphics[angle=0,width=160mm]{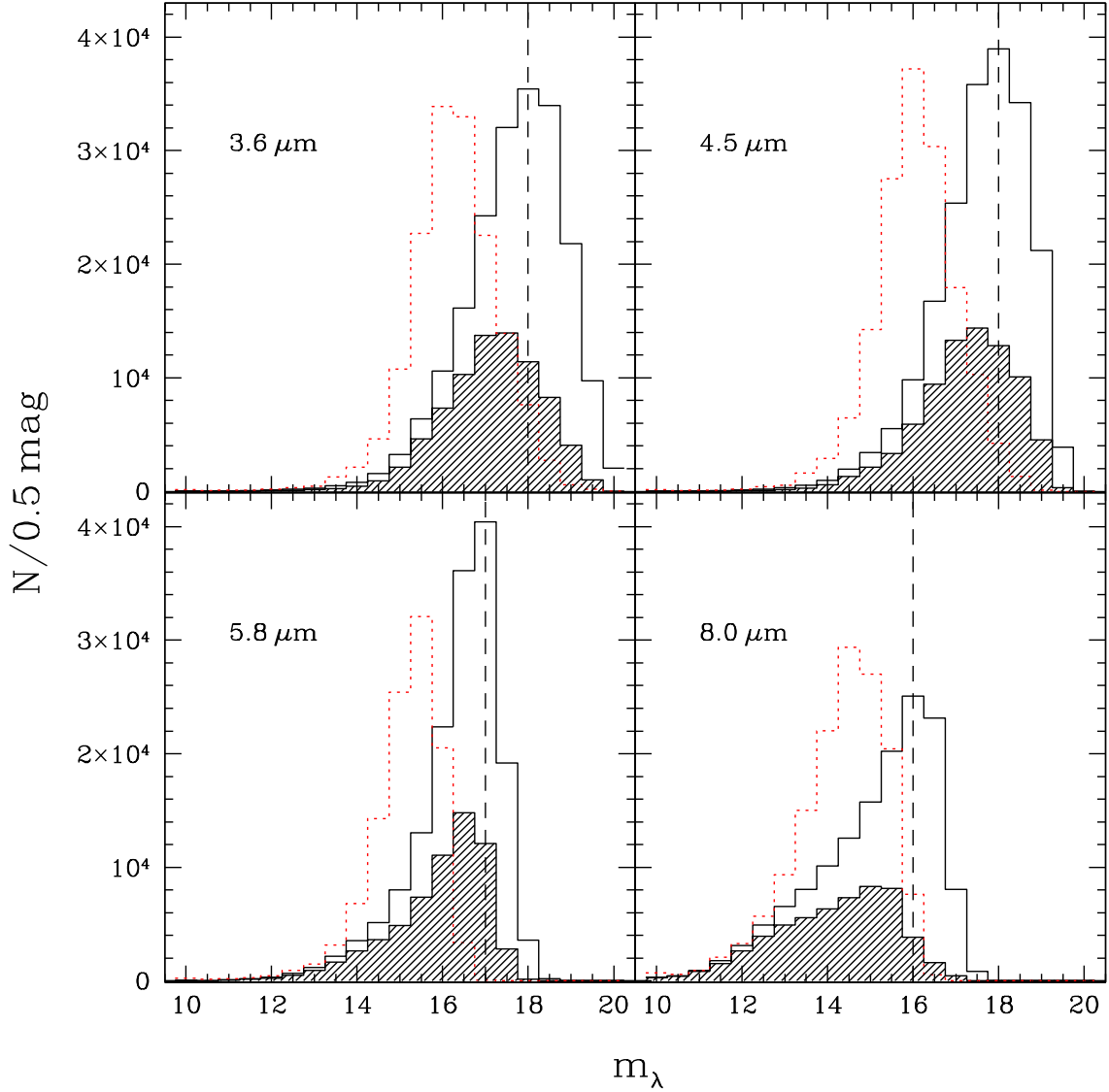}
\end{center}
\caption{Apparent magnitude histograms for all the sources in the catalog, 
with the shaded regions showing the sources in M\,33.
Overall, the source lists become incomplete at (locations marked by the dashed vertical
lines) $m_{3.6}\gtrsim18$, $m_{4.5}\gtrsim18$, $m_{5.8}\gtrsim17$ and $m_{8.0}\gtrsim16$.
The M\,33 source list is roughly $\sim0.5$\,mag shallower than 
for the other galaxies. The dotted lines show the apparent-magnitude 
histograms of the SDWFS catalog sources, scaled up  for clarity by a factor of $7$
for $m_{3.6}$ and $m_{4.5}$, and by a factor of $10$ for $m_{5.8}$ and $m_{8.0}$.}
\label{fig:mHist}
\end{figure}

\begin{figure}
\begin{center}
\includegraphics[angle=0,width=160mm]{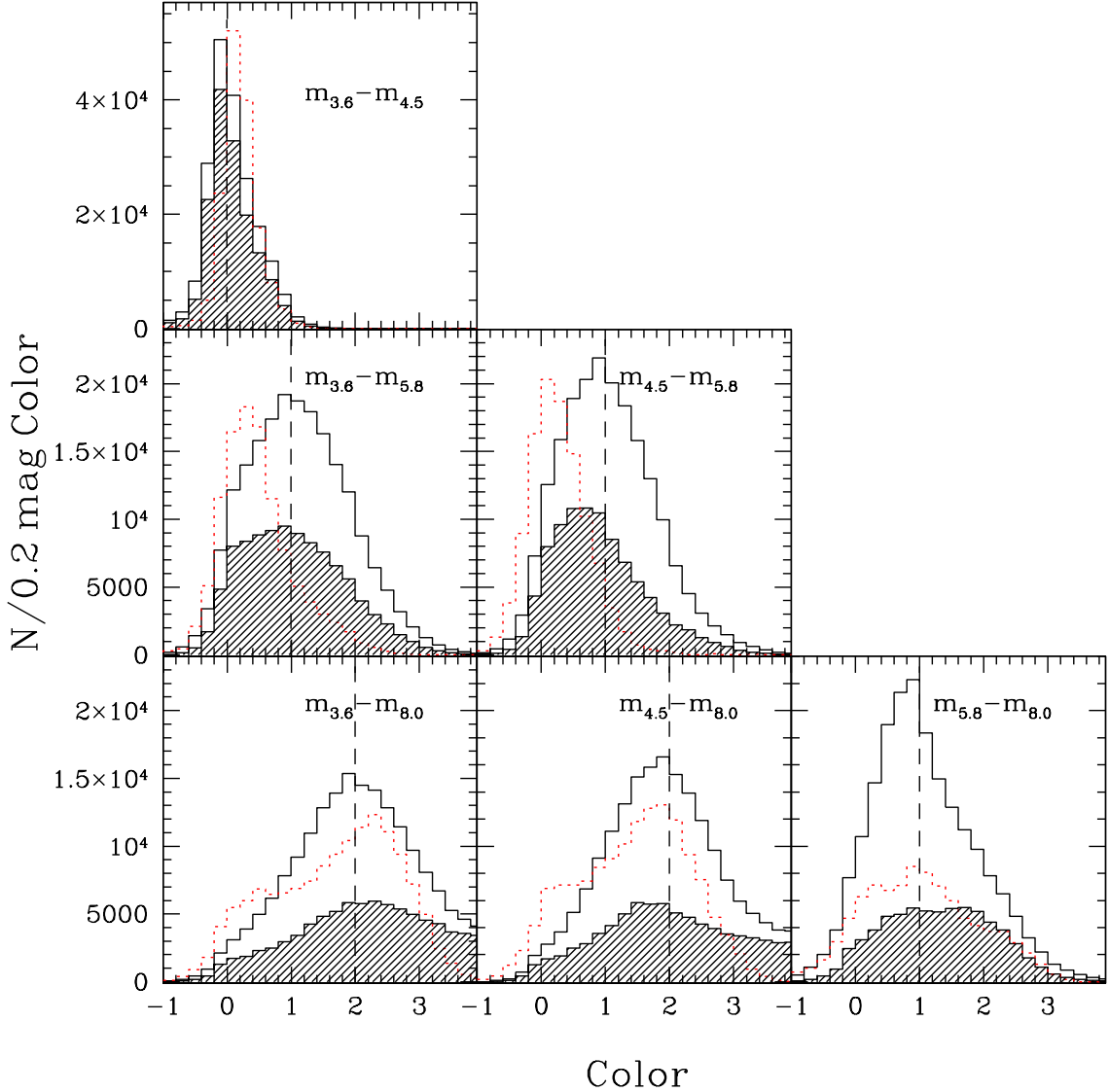}
\end{center}
\caption{Mid-IR color histograms for all the sources in the catalog, 
with the shaded histograms showing the distribution of sources with 
$1\,\sigma$ uncertainty in color $\lesssim0.2$.
The sharp peak of color distribution at $m_{3.6}-m_{4.5}\simeq0$ is due to the 
Rayleigh-Jeans tail of the spectra of all normal stars.
The longer wavelengths are increasingly dominated by dusty 
stars, with color distribution peaks at (locations marked by the dashed vertical
lines)
$m_{3.6}-m_{5.8}=m_{4.5}-m_{5.8}\simeq1$, $m_{3.6}-m_{8.0}=m_{4.5}-m_{8.0}\simeq2$,
and $m_{5.8}-m_{8.0}\simeq1$. The dotted lines show the mid-IR color
histograms of the SDWFS catalog sources, scaled-up  for clarity by a factor of $7$
for $m_{3.6}-m_{4.5}$, and a factor of $10$ for the rest..}
\label{fig:cHist}
\end{figure}

Figure \ref{fig:mHist} 
show the apparent magnitude histograms of all sources in the catalog, 
with the shaded regions showing the sources in M\,33.
It is apparent that the M\,33 catalog 
is $\sim0.5$\,mag shallower than the other catalogs in all bands. 
Overall, our source lists become significantly incomplete at $m_{3.6}\gtrsim18$, $m_{4.5}\gtrsim18$, 
$m_{5.8}\gtrsim17$ and $m_{8.0}\gtrsim16$. Figure\,\ref{fig:cHist} 
show the mid-IR color histograms of all sources in the catalog, 
with the shaded regions showing the sources with
$1\,\sigma$ uncertainty in color $\lesssim0.2$.
It demonstrates the absence of any surprises in the color distributions and 
that the mid-color distributions are not due to large uncertainties.
All normal stars have the same mid-IR color in the first two IRAC bands,
because of the Rayleigh-Jeans
tails of their spectra, and we see this from the sharp peak of
color distribution at $m_{3.6}-m_{4.5}\simeq0$. 
The longer wavelength detections are increasingly dominated by dusty stars, 
with color distribution peaks at 
$m_{3.6}-m_{5.8}=m_{4.5}-m_{5.8}\simeq1$, $m_{3.6}-m_{8.0}=m_{4.5}-m_{8.0}\simeq2$,
and $m_{5.8}-m_{8.0}\simeq1$. Figures \ref{fig:mHist} and \ref{fig:cHist}
also show the apparent magnitude and color histograms of the SDWFS point 
sources (dotted lines). They show that our catalogs are $>1$\,mag deeper than the 
SDWFS catalog and the mid-IR color distribution of 
blank-field extragalactic sources significantly differs from that
of fields containing nearby galaxies. In particular, at longer wavelengths,
the galaxies contain more red sources (dusty stars) while the 
random extragalactic field sources are generally bluer.

Although we report the $24\,\micron$ photometry in the catalogs,
these measurements have limited utility due to the lower spatial 
resolution of this band.
The aperture used for this band commonly includes 
objects other than the intended target and is contaminated 
by emission from cold interstellar dust. Nevertheless,
as we showed in Figure\,5 of \citet{ref:Khan_2013}, the spectral energy 
distributions of the normal stars show the expected negative slope for the Rayleigh-Jeans tail 
of their SEDs between $8\micron$ and 
$24\,\micron$.
The $24\,\micron$ photometry can be very useful in specific cases, such as 
for studying evolved massive stars \citep[see][for details]{ref:Khan_2015},
despite the resolution limitation.

Where \citet{ref:Thompson_2009} identified $\sim53,200$ sources
in M\,33, and \citet{ref:Khan_2010} identified $\sim11,200$ sources 
in NGC\,300 and $\sim6,000$ sources in M\,81, now we catalog 
$\sim78,800$, $\sim21,700$ and $\sim28,600$ sources in these galaxies 
respectively. This is due to a number of factors. First, both of the  
earlier studies only cataloged the central regions of these 
galaxies, while here we analyze the full mosaics. 
Second, we use a larger matching radius 
of $1$\,pixel (rather than $0.5$\,pixel) to define point sources and
a higher fraction of our cataloged sources 
can be coincidental matches between the $3.6\,\micron$ and $4.5\,\micron$. 
Third, the catalogs presented here 
are deeper due to improved photometry and search methods based on 
lessons learned from \citet{ref:Khan_2011,ref:Khan_2013,ref:Khan_2015}. For example, \citet{ref:Thompson_2009} 
noted that their M\,$33$ point-source list becomes incomplete at $m_{3.6}=m_{4.5}>17.1$,
while we reach $\sim0.5$\,mag deeper for this galaxy and $\sim1$\,mag 
deeper for the rest. 

\begin{figure}
\begin{center}
\includegraphics[angle=0,width=160mm]{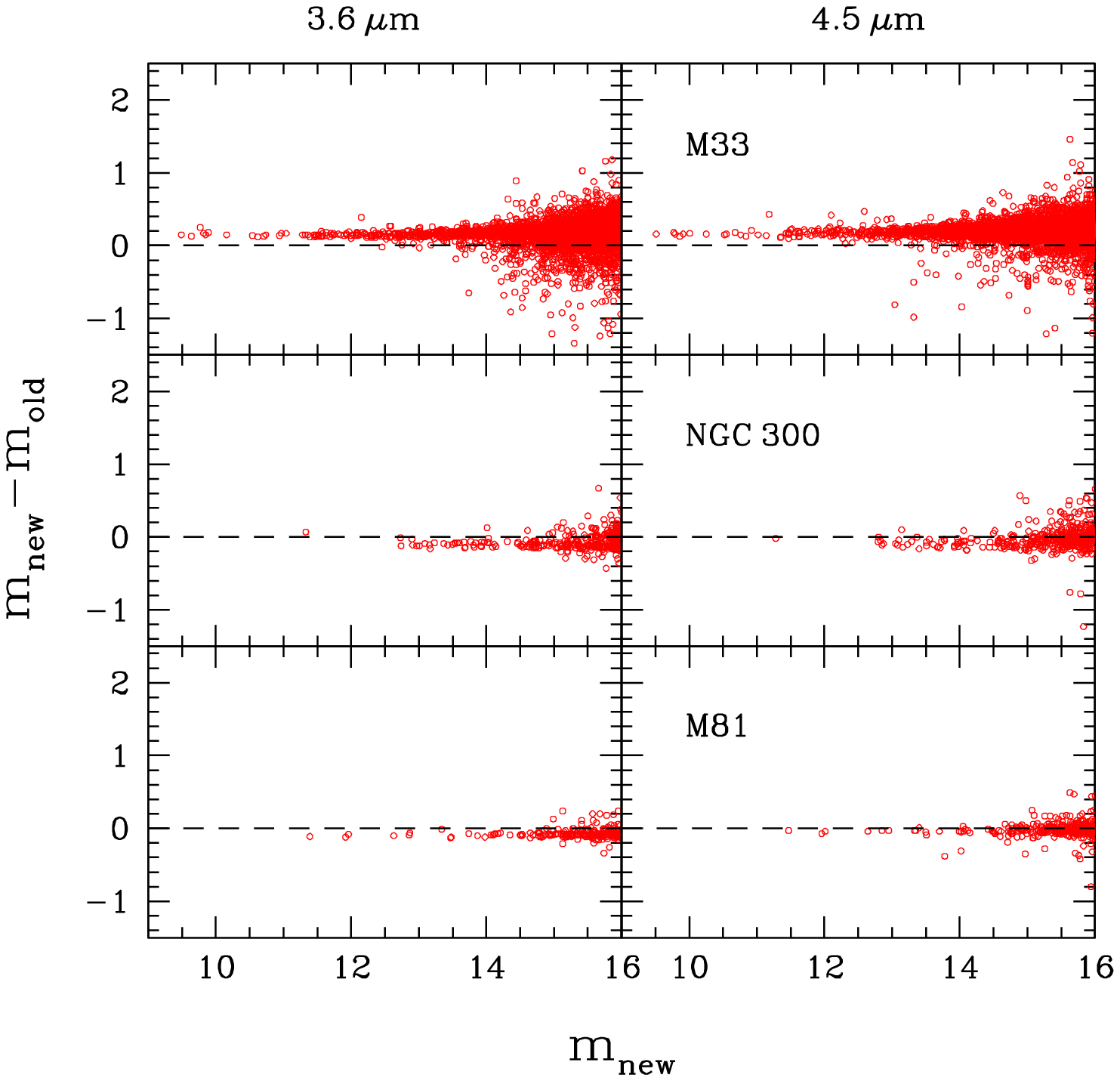}
\end{center}
\caption{The differences between $3.6\,\micron$ and $4.5\,\micron$
magnitudes reported by \citet{ref:Thompson_2009} for M\,33 and by
\citet{ref:Khan_2010} for NGC\,300 and M\,81 ($m_{old}$) as compared to
our measurements at those bands for the same sources ($m_{new}$). 
Aside from the linear offsets due to calibration differences, as discussed in 
Section\,\ref{sec:cats}, the photometric measurements at 
these two bands agree 
within the stated uncertainties for the brighter sources, with the 
scatter increasing towards the fainter sources.}
\label{fig:compare}
\end{figure}

We have compared our M\,33, NGC\,300 and M\,81 
catalogs with the  $3.6\,\micron$ and $4.5\,\micron$ 
catalogs published by \citet{ref:Thompson_2009} and 
\citet{ref:Khan_2010}. The photometric measurements at 
these two bands agree within the stated uncertainties 
for the brighter sources, with the scatter increasing 
towards the fainter sources. As is apparent from 
Figure\,\ref{fig:compare}, we identify zero point linear offsets 
between the older catalogs and the measurements reported 
here due to calibration differences. \citet{ref:Thompson_2009} 
performed PSF-fitting photometry and 
then converted the measurements to Vega-calibrated magnitudes 
using simple flux zero point shifts derived from aperture photometry 
flux measurements of $10-20$ bright and isolated stars in each 
band. On the other hand, in this work as well as in 
\citet{ref:Khan_2010}, we perform the PSF and aperture 
photometry measurements independently and then convert both to 
Vega-calibrated magnitudes using the flux zero points provided 
in the \textit{Spitzer} Data Analysis Cookbook. This appears to 
have led to the $\sim0.2\,$magnitude linear zero-point offset between the 
\citet{ref:Thompson_2009} calibrations and our measurements for M\,33.

The more subtle offset between the \citet{ref:Khan_2010} 
calibrations and the measurements we report here for NGC\,300 and M\,81 are due to the 
earlier work having used a background annulus far from the aperture 
($1\farcs5$ aperture, $9\farcs-15\farcs$ annulus)
to measure the local sky brightness when performing aperture photometry, 
whereas in this work we use a background annulus immediately adjacent to 
the aperture ($2\farcs4$ aperture, $2\farcs4-7\farcs2$ annulus). In a crowded 
field, where a source is likely to be contaminated by flux from adjacent 
sources, an annulus far from the aperture can underestimate the local sky.
As the aperture photometry measurement is used to fine tune the PSF photometry 
to account for PSF fitting up to a finite radius rather than to infinity,
this appears to have caused the $<0.1\,$magnitude offset. We identified this 
issue previously and adopted the current practice of using a background 
annulus immediately adjacent to the aperture \citep{ref:Khan_2013,ref:Khan_2015}.

Point-source catalogs of the inherently crowded galaxy fields that we are surveying are 
bound to be crowding limited, not just magnitude limited. 
While Figure\,\ref{fig:cHist} empirically demonstrates that our source detection 
peaks at a certain magnitude and then falls off rapidly, it is likely that 
incompleteness is affecting even the bright-star counts, increasing towards and 
through the peak. For example, \citet{ref:Martinez_2015} suggests that even in 
a sparse field, incompleteness at the $5\sigma$ detection level can be on the 
order of $15\%$. Ideally, performing an artificial star completeness 
test through addition of randomly distributed artificial objects in the
images could be useful.

However, the mid-IR bright stars (massive stars) are 
not randomly distributed, but in fact are highly clustered. 
Performing an efficiency determination test through addition of randomly distributed 
artificial objects in the images therefore would lead us to either overestimate or
underestimate the efficiency. For such a study to be truly useful, it would require a 
proper ``star-star correlation function'' to be employed for spatial distribution of 
artificial stars. A star-star correlation function can also characterize the typical 
scale of the star clusters in each galaxy as a function of magnitude, to investigate 
the possible affect of aliasing. Such detailed modeling of source distribution 
is beyond the scope of the current paper and we encourage future studies to 
explore this issue.

This catalog is a resource as an archive for studying mid-IR transients
and for planning observations with the James Webb Space Telescope.
Our survey is being expanded to 
galaxies with $\sim10$x higher integrated star formation rate 
than for these seven galaxies.
While we have shown that surveys for stellar populations are feasible
using archival Spitzer data, JWST will be a far more powerful probe
of stars in the mid-IR. The nearly order-of-magnitude higher resolution 
\citep{ref:Gardner_2006} of JWST compared to {\it Spitzer}
can be used either to greatly reduce the problem of confusion 
in these galaxies or to greatly expand the survey volume. 

\acknowledgments
We would like to thank the referee for providing helpful feedback.
This work is based on observations made with the {\it Spitzer} Space Telescope, 
which is operated by the Jet Propulsion Laboratory, California Institute of Technology 
under a contract with the National Aeronautics and Space Administration (NASA). 
We extend our gratitude to the SINGS Legacy Survey 
and the LVL Survey for making their data publicly available. 
RK is supported through a JWST Fellowship hosted by the Goddard Space Flight Center and 
awarded as part of the NASA Postdoctoral Program operated by the Oak Ridge Associated 
Universities on behalf of NASA.

\clearpage

\clearpage

\clearpage

\clearpage

\clearpage

\clearpage

\clearpage

\clearpage

\clearpage

\clearpage

\begin{appendix}
\label{appendix}

\begin{landscape}

\begin{table}
\begin{center}
\begin{small} 
\caption{Catalog Statistics}
\label{tab:stats}
\begin{tabular}{lrrrrrrr}
\\
\hline 
\hline
\\
\multicolumn{1}{c}{} &
\multicolumn{1}{c}{NGC} &
\multicolumn{1}{c}{M\,$33$} &
\multicolumn{1}{c}{NGC} &
\multicolumn{1}{c}{NGC} &
\multicolumn{1}{c}{M\,$81$} &
\multicolumn{1}{c}{NGC} &
\multicolumn{1}{c}{NGC} 
\\
\multicolumn{1}{c}{} &
\multicolumn{1}{c}{$6822$} &
\multicolumn{1}{c}{} &
\multicolumn{1}{c}{$300$} &
\multicolumn{1}{c}{$2403$} &
\multicolumn{1}{c}{} &
\multicolumn{1}{c}{$247$} &
\multicolumn{1}{c}{$7793$} 
\\
\\
\hline 
\hline
\\
Survey Area (deg$^2$)  & $\sim0.1$    & $\sim0.73$   & $\sim0.17$   & $\sim0.12$   & $\sim0.17$   & $\sim0.2$    & $\sim0.044$ \\

SFR ($M_\odot / year$)            & $0.01$   & $0.33$   & $0.11$   & $0.44$   & $0.46$   & $0.17$   &  $0.33$ \\

Point Sources                & $30,745$ & $78,841$ & $21,739$ & $17,022$ & $28,674$ & $16,658$ & $5,617$ \\

Matched within $<0.5$\,pixel  & $24,134$ & $71,439$ & $16,209$ & $12,563$ & $18,383$ & $12,741$ & $3,401$ \\

Matched within $0.5-1.0$\,pixel & $6,611$ & $7,402$ & $5,530$ & $4,459$ & $10,291$ & $3,917$ & $2,216$ \\

Duplicate Matches  & $142$    &    $271$ &     $64$ &     $87$ &     $66$ &     $39$ &    $54$ \\

$>3\sigma$ at $5.8\,\micron$ & $23,906$ & $62,974$ & $18,043$ & $12,712$ & $21,332$ & $13,893$ & $3,389$ \\

$>3\sigma$ at $8.0\,\micron$ & $20,414$ & $56,145$ & $17,204$ & $12,113$ & $20,069$ & $11,995$ & $4,018$ \\

$>3\sigma$ at $24\,\micron$  &  $7,268$ & $23,004$ &  $7,322$ &  $6,886$ & $11,557$ &  $5,297$ & $2,478$ \\
\\
\hline
\hline
\end{tabular}
\end{small} 
\end{center}
\end{table}

\begin{table}   
\begin{center}   
\begin{small}  
\caption{Catalog for $30,745$ Point Sources in NGC\,$6822$} 
\label{tab:n6822}   
\begin{tabular}{rrrrrrrrrrrrrrrrr}   
\\   
\hline  
\hline  
\multicolumn{1}{c}{RA} &
\multicolumn{1}{c}{Dec} &
\multicolumn{1}{c}{} &
\multicolumn{1}{c}{$m_{3.6}$} &
\multicolumn{1}{c}{$\sigma_{3.6}$} &
\multicolumn{1}{c}{$\delta_{3.6}$} &
\multicolumn{1}{c}{$m_{4.5}$} &
\multicolumn{1}{c}{$\sigma_{4.5}$} &
\multicolumn{1}{c}{$\delta_{4.5}$} &
\multicolumn{1}{c}{$m_{5.8}$} &
\multicolumn{1}{c}{$\sigma_{5.8}$} &
\multicolumn{1}{c}{$\delta_{5.8}$} &
\multicolumn{1}{c}{$m_{8.0}$} &
\multicolumn{1}{c}{$\sigma_{8.0}$} &
\multicolumn{1}{c}{$\delta_{8.0}$} &
\multicolumn{1}{c}{$m_{24}$} &
\multicolumn{1}{c}{$\sigma_{24}$} 
\\
\multicolumn{1}{c}{(deg)} &
\multicolumn{1}{c}{(deg)} &
\multicolumn{1}{c}{} &
\multicolumn{1}{c}{(mag)} &
\multicolumn{1}{c}{} &
\multicolumn{1}{c}{} &
\multicolumn{1}{c}{(mag)} &
\multicolumn{1}{c}{} &
\multicolumn{1}{c}{} &
\multicolumn{1}{c}{(mag)} &
\multicolumn{1}{c}{} &
\multicolumn{1}{c}{} &
\multicolumn{1}{c}{(mag)} &
\multicolumn{1}{c}{} &
\multicolumn{1}{c}{} &
\multicolumn{1}{c}{(mag)} &
\multicolumn{1}{c}{} 
\\ 
\hline  
\hline
\dots        & \dots       && \dots  & \dots  & \dots  & \dots  & \dots  & \dots  & \dots  & \dots  & \dots  & \dots  & \dots  & \dots  & \dots  & \dots  \\
 $ 296.37946 $ & $ -14.74545 $ & & $  9.07 $ & $ 0.12 $ & $ -0.22 $ & $  8.94 $ & $ 0.10 $ & $ -0.09 $ & $  9.00 $ & $ 0.03 $ & $  0.02 $ & $  9.34 $ & $ 0.04 $ & $  0.21 $ & $  8.80 $ & $  0.01 $  \\
 $ 296.23211 $ & $ -14.72759 $ & & $  9.34 $ & $ 0.11 $ & $ -0.18 $ & $  9.09 $ & $ 0.03 $ & $ -0.08 $ & $  9.14 $ & $ 0.02 $ & $  0.00 $ & $  9.24 $ & $ 0.04 $ & $  0.05 $ & $  9.13 $ & $  0.02 $  \\
 $ 296.31297 $ & $ -14.77366 $ & & $  9.48 $ & $ 0.06 $ & $ -0.07 $ & $  9.35 $ & $ 0.05 $ & $  0.10 $ & $  9.16 $ & $ 0.04 $ & $ -0.01 $ & $  9.34 $ & $ 0.01 $ & $  0.12 $ & $  9.08 $ & $  0.03 $  \\
 $ 296.32372 $ & $ -14.71925 $ & & $  9.46 $ & $ 0.03 $ & $ -0.10 $ & $  9.37 $ & $ 0.04 $ & $  0.03 $ & $  9.33 $ & $ 0.03 $ & $ -0.01 $ & $  9.59 $ & $ 0.04 $ & $  0.13 $ & $  9.30 $ & $  0.02 $  \\
 $ 296.13722 $ & $ -14.94461 $ & & $  9.77 $ & $ 0.06 $ & $ -0.01 $ & $  9.39 $ & $ 0.07 $ & $ -0.16 $ & $  9.54 $ & $ 0.03 $ & $  0.00 $ & $  9.72 $ & $ 0.02 $ & $  0.12 $ & $  9.37 $ & $  0.02 $  \\
\dots        & \dots       && \dots  & \dots  & \dots  & \dots  & \dots  & \dots  & \dots  & \dots  & \dots  & \dots  & \dots  & \dots  & \dots  & \dots  \\
\hline  
\hline  
\end{tabular} 
\end{small} 
\end{center}  
\end{table}

\end{landscape}

\clearpage

\begin{landscape}

\begin{table}   
\begin{center}   
\begin{small}  
\caption{Catalog for $78,841$ Point Sources in M\,$33$} 
\label{tab:m33}   
\begin{tabular}{rrrrrrrrrrrrrrrrr}   
\\   
\hline  
\hline  
\multicolumn{1}{c}{RA} &
\multicolumn{1}{c}{Dec} &
\multicolumn{1}{c}{} &
\multicolumn{1}{c}{$m_{3.6}$} &
\multicolumn{1}{c}{$\sigma_{3.6}$} &
\multicolumn{1}{c}{$\delta_{3.6}$} &
\multicolumn{1}{c}{$m_{4.5}$} &
\multicolumn{1}{c}{$\sigma_{4.5}$} &
\multicolumn{1}{c}{$\delta_{4.5}$} &
\multicolumn{1}{c}{$m_{5.8}$} &
\multicolumn{1}{c}{$\sigma_{5.8}$} &
\multicolumn{1}{c}{$\delta_{5.8}$} &
\multicolumn{1}{c}{$m_{8.0}$} &
\multicolumn{1}{c}{$\sigma_{8.0}$} &
\multicolumn{1}{c}{$\delta_{8.0}$} &
\multicolumn{1}{c}{$m_{24}$} &
\multicolumn{1}{c}{$\sigma_{24}$} 
\\
\multicolumn{1}{c}{(deg)} &
\multicolumn{1}{c}{(deg)} &
\multicolumn{1}{c}{} &
\multicolumn{1}{c}{(mag)} &
\multicolumn{1}{c}{} &
\multicolumn{1}{c}{} &
\multicolumn{1}{c}{(mag)} &
\multicolumn{1}{c}{} &
\multicolumn{1}{c}{} &
\multicolumn{1}{c}{(mag)} &
\multicolumn{1}{c}{} &
\multicolumn{1}{c}{} &
\multicolumn{1}{c}{(mag)} &
\multicolumn{1}{c}{} &
\multicolumn{1}{c}{} &
\multicolumn{1}{c}{(mag)} &
\multicolumn{1}{c}{} 
\\ 
\hline  
\hline
\dots        & \dots       && \dots  & \dots  & \dots  & \dots  & \dots  & \dots  & \dots  & \dots  & \dots  & \dots  & \dots  & \dots  & \dots  & \dots  \\
 $  23.40441 $ & $  30.38922 $ & & $  7.72 $ & $ 0.05 $ & $ -0.69 $ & $  7.75 $ & $ 0.03 $ & $ -0.20 $ & $  7.66 $ & $ 0.02 $ & $  0.04 $ & $  7.60 $ & $ 0.06 $ & $  0.01 $ & $  7.65 $ & $  0.05 $  \\
 $  23.66245 $ & $  30.84435 $ & & $  7.86 $ & $ 0.03 $ & $ -0.60 $ & $  7.89 $ & $ 0.02 $ & $ -0.15 $ & $  7.74 $ & $ 0.01 $ & $  0.02 $ & $  7.65 $ & $ 0.11 $ & $  0.01 $ & $  7.69 $ & $  0.01 $  \\
 $  23.82179 $ & $  30.67099 $ & & $  8.14 $ & $ 0.04 $ & $ -0.49 $ & $  8.20 $ & $ 0.02 $ & $ -0.10 $ & $  8.13 $ & $ 0.01 $ & $  0.05 $ & $  8.02 $ & $ 0.10 $ & $ -0.04 $ & $  8.13 $ & $  0.01 $  \\
 $  23.23681 $ & $  30.26192 $ & & $  7.99 $ & $ 0.04 $ & $ -0.85 $ & $  8.54 $ & $ 0.03 $ & $ -0.20 $ & $  8.03 $ & $ 0.02 $ & $  0.04 $ & $  8.25 $ & $ 0.10 $ & $  0.04 $ & $  7.51 $ & $  0.01 $  \\
 $  23.34689 $ & $  30.94818 $ & & $  8.54 $ & $ 0.03 $ & $ -0.24 $ & $  9.25 $ & $ 0.03 $ & $  0.04 $ & $  8.39 $ & $ 0.01 $ & $  0.03 $ & $  8.97 $ & $ 0.08 $ & $ -0.15 $ & $  8.40 $ & $  0.01 $  \\
\dots        & \dots       && \dots  & \dots  & \dots  & \dots  & \dots  & \dots  & \dots  & \dots  & \dots  & \dots  & \dots  & \dots  & \dots  & \dots  \\
\hline  
\hline  
\end{tabular} 
\end{small} 
\end{center}  
\end{table}

\begin{table}   
\begin{center}   
\begin{small}  
\caption{Catalog for $21,739$ Point Sources in NGC\,$300$} 
\label{tab:n300}   
\begin{tabular}{rrrrrrrrrrrrrrrrr}   
\\   
\hline  
\hline  
\multicolumn{1}{c}{RA} &
\multicolumn{1}{c}{Dec} &
\multicolumn{1}{c}{} &
\multicolumn{1}{c}{$m_{3.6}$} &
\multicolumn{1}{c}{$\sigma_{3.6}$} &
\multicolumn{1}{c}{$\delta_{3.6}$} &
\multicolumn{1}{c}{$m_{4.5}$} &
\multicolumn{1}{c}{$\sigma_{4.5}$} &
\multicolumn{1}{c}{$\delta_{4.5}$} &
\multicolumn{1}{c}{$m_{5.8}$} &
\multicolumn{1}{c}{$\sigma_{5.8}$} &
\multicolumn{1}{c}{$\delta_{5.8}$} &
\multicolumn{1}{c}{$m_{8.0}$} &
\multicolumn{1}{c}{$\sigma_{8.0}$} &
\multicolumn{1}{c}{$\delta_{8.0}$} &
\multicolumn{1}{c}{$m_{24}$} &
\multicolumn{1}{c}{$\sigma_{24}$} 
\\
\multicolumn{1}{c}{(deg)} &
\multicolumn{1}{c}{(deg)} &
\multicolumn{1}{c}{} &
\multicolumn{1}{c}{(mag)} &
\multicolumn{1}{c}{} &
\multicolumn{1}{c}{} &
\multicolumn{1}{c}{(mag)} &
\multicolumn{1}{c}{} &
\multicolumn{1}{c}{} &
\multicolumn{1}{c}{(mag)} &
\multicolumn{1}{c}{} &
\multicolumn{1}{c}{} &
\multicolumn{1}{c}{(mag)} &
\multicolumn{1}{c}{} &
\multicolumn{1}{c}{} &
\multicolumn{1}{c}{(mag)} &
\multicolumn{1}{c}{} 
\\ 
\hline  
\hline
\dots        & \dots       && \dots  & \dots  & \dots  & \dots  & \dots  & \dots  & \dots  & \dots  & \dots  & \dots  & \dots  & \dots  & \dots  & \dots  \\
 $  13.45595 $ & $ -37.68470 $ & & $ 10.96 $ & $ 0.09 $ & $  0.72 $ & $ 10.13 $ & $ 0.04 $ & $ -0.03 $ & $ 10.19 $ & $ 0.01 $ & $  0.04 $ & $ 10.20 $ & $ 0.01 $ & $  0.11 $ & $ 10.11 $ & $  0.03 $  \\
 $  13.87874 $ & $ -37.65598 $ & & $ 10.23 $ & $ 0.09 $ & $ -0.05 $ & $ 10.31 $ & $ 0.04 $ & $  0.08 $ & $ 10.19 $ & $ 0.03 $ & $  0.02 $ & $ 10.28 $ & $ 0.02 $ & $  0.08 $ & $  9.97 $ & $  0.02 $  \\
 $  13.62935 $ & $ -37.77459 $ & & $ 10.43 $ & $ 0.04 $ & $ -0.11 $ & $ 10.54 $ & $ 0.04 $ & $  0.01 $ & $ 10.57 $ & $ 0.01 $ & $  0.03 $ & $ 10.62 $ & $ 0.02 $ & $  0.10 $ & $ 10.58 $ & $  0.05 $  \\
 $  13.74223 $ & $ -37.51812 $ & & $ 10.91 $ & $ 0.04 $ & $ -0.06 $ & $ 10.99 $ & $ 0.02 $ & $ -0.01 $ & $ 10.92 $ & $ 0.02 $ & $  0.01 $ & $ 11.03 $ & $ 0.02 $ & $  0.08 $ & $ 11.30 $ & $  0.12 $  \\
 $  13.92891 $ & $ -37.74328 $ & & $ 11.09 $ & $ 0.05 $ & $  0.02 $ & $ 11.11 $ & $ 0.03 $ & $  0.01 $ & $ 11.06 $ & $ 0.02 $ & $  0.02 $ & $ 11.18 $ & $ 0.02 $ & $  0.09 $ & $ 11.03 $ & $  0.10 $  \\
\dots        & \dots       && \dots  & \dots  & \dots  & \dots  & \dots  & \dots  & \dots  & \dots  & \dots  & \dots  & \dots  & \dots  & \dots  & \dots  \\
\hline  
\hline  
\end{tabular} 
\end{small} 
\end{center}  
\end{table}

\end{landscape}

\clearpage

\begin{landscape}

\begin{table}   
\begin{center}   
\begin{small}  
\caption{Catalog for $17,022$ Point Sources in NGC\,$2403$} 
\label{tab:n2403}   
\begin{tabular}{rrrrrrrrrrrrrrrrr}   
\\   
\hline  
\hline  
\multicolumn{1}{c}{RA} &
\multicolumn{1}{c}{Dec} &
\multicolumn{1}{c}{} &
\multicolumn{1}{c}{$m_{3.6}$} &
\multicolumn{1}{c}{$\sigma_{3.6}$} &
\multicolumn{1}{c}{$\delta_{3.6}$} &
\multicolumn{1}{c}{$m_{4.5}$} &
\multicolumn{1}{c}{$\sigma_{4.5}$} &
\multicolumn{1}{c}{$\delta_{4.5}$} &
\multicolumn{1}{c}{$m_{5.8}$} &
\multicolumn{1}{c}{$\sigma_{5.8}$} &
\multicolumn{1}{c}{$\delta_{5.8}$} &
\multicolumn{1}{c}{$m_{8.0}$} &
\multicolumn{1}{c}{$\sigma_{8.0}$} &
\multicolumn{1}{c}{$\delta_{8.0}$} &
\multicolumn{1}{c}{$m_{24}$} &
\multicolumn{1}{c}{$\sigma_{24}$} 
\\
\multicolumn{1}{c}{(deg)} &
\multicolumn{1}{c}{(deg)} &
\multicolumn{1}{c}{} &
\multicolumn{1}{c}{(mag)} &
\multicolumn{1}{c}{} &
\multicolumn{1}{c}{} &
\multicolumn{1}{c}{(mag)} &
\multicolumn{1}{c}{} &
\multicolumn{1}{c}{} &
\multicolumn{1}{c}{(mag)} &
\multicolumn{1}{c}{} &
\multicolumn{1}{c}{} &
\multicolumn{1}{c}{(mag)} &
\multicolumn{1}{c}{} &
\multicolumn{1}{c}{} &
\multicolumn{1}{c}{(mag)} &
\multicolumn{1}{c}{} 
\\ 
\hline  
\hline
\dots        & \dots       && \dots  & \dots  & \dots  & \dots  & \dots  & \dots  & \dots  & \dots  & \dots  & \dots  & \dots  & \dots  & \dots  & \dots  \\
 $ 114.26503 $ & $  65.49629 $ & & $  9.88 $ & $ 0.04 $ & $ -0.06 $ & $  9.86 $ & $ 0.01 $ & $ -0.06 $ & $  9.89 $ & $ 0.02 $ & $ -0.00 $ & $ 10.15 $ & $ 0.04 $ & $  0.22 $ & $  9.90 $ & $  0.03 $  \\
 $ 113.70299 $ & $  65.56170 $ & & $  9.92 $ & $ 0.06 $ & $ -0.09 $ & $ 10.00 $ & $ 0.02 $ & $ -0.06 $ & $ 10.01 $ & $ 0.02 $ & $  0.01 $ & $  9.88 $ & $ 0.06 $ & $ -0.18 $ & $  9.93 $ & $  0.05 $  \\
 $ 114.20327 $ & $  65.59495 $ & & $ 10.17 $ & $ 0.06 $ & $ -0.04 $ & $ 10.09 $ & $ 0.05 $ & $ -0.07 $ & $ 10.18 $ & $ 0.05 $ & $ -0.00 $ & $ 10.36 $ & $ 0.04 $ & $ -0.08 $ & $  8.87 $ & $  0.20 $  \\
 $ 114.04298 $ & $  65.75131 $ & & $ 10.43 $ & $ 0.07 $ & $  0.00 $ & $ 10.41 $ & $ 0.08 $ & $ -0.09 $ & $ 10.44 $ & $ 0.03 $ & $ -0.02 $ & $ 10.27 $ & $ 0.06 $ & $ -0.18 $ & $ 10.11 $ & $  0.04 $  \\
 $ 113.65425 $ & $  65.53209 $ & & $ 10.93 $ & $ 0.15 $ & $  0.14 $ & $ 10.42 $ & $ 0.11 $ & $ -0.07 $ & $ 10.86 $ & $ 0.14 $ & $  0.18 $ & $ 10.44 $ & $ 0.04 $ & $ -0.05 $ & $ 10.50 $ & $  0.04 $  \\
\dots        & \dots       && \dots  & \dots  & \dots  & \dots  & \dots  & \dots  & \dots  & \dots  & \dots  & \dots  & \dots  & \dots  & \dots  & \dots  \\
\hline  
\hline  
\end{tabular} 
\end{small} 
\end{center}  
\end{table}

\begin{table}   
\begin{center}   
\begin{small}  
\caption{Catalog for $28,674$ Point Sources in M\,$81$} 
\label{tab:m81}   
\begin{tabular}{rrrrrrrrrrrrrrrrr}   
\\   
\hline  
\hline  
\multicolumn{1}{c}{RA} &
\multicolumn{1}{c}{Dec} &
\multicolumn{1}{c}{} &
\multicolumn{1}{c}{$m_{3.6}$} &
\multicolumn{1}{c}{$\sigma_{3.6}$} &
\multicolumn{1}{c}{$\delta_{3.6}$} &
\multicolumn{1}{c}{$m_{4.5}$} &
\multicolumn{1}{c}{$\sigma_{4.5}$} &
\multicolumn{1}{c}{$\delta_{4.5}$} &
\multicolumn{1}{c}{$m_{5.8}$} &
\multicolumn{1}{c}{$\sigma_{5.8}$} &
\multicolumn{1}{c}{$\delta_{5.8}$} &
\multicolumn{1}{c}{$m_{8.0}$} &
\multicolumn{1}{c}{$\sigma_{8.0}$} &
\multicolumn{1}{c}{$\delta_{8.0}$} &
\multicolumn{1}{c}{$m_{24}$} &
\multicolumn{1}{c}{$\sigma_{24}$} 
\\
\multicolumn{1}{c}{(deg)} &
\multicolumn{1}{c}{(deg)} &
\multicolumn{1}{c}{} &
\multicolumn{1}{c}{(mag)} &
\multicolumn{1}{c}{} &
\multicolumn{1}{c}{} &
\multicolumn{1}{c}{(mag)} &
\multicolumn{1}{c}{} &
\multicolumn{1}{c}{} &
\multicolumn{1}{c}{(mag)} &
\multicolumn{1}{c}{} &
\multicolumn{1}{c}{} &
\multicolumn{1}{c}{(mag)} &
\multicolumn{1}{c}{} &
\multicolumn{1}{c}{} &
\multicolumn{1}{c}{(mag)} &
\multicolumn{1}{c}{} 
\\ 
\hline  
\hline
\dots        & \dots       && \dots  & \dots  & \dots  & \dots  & \dots  & \dots  & \dots  & \dots  & \dots  & \dots  & \dots  & \dots  & \dots  & \dots  \\
 $ 148.36619 $ & $  68.97859 $ & & $  9.38 $ & $ 0.06 $ & $ -0.23 $ & $  9.30 $ & $ 0.10 $ & $ -0.03 $ & $  9.33 $ & $ 0.02 $ & $ -0.01 $ & $  9.52 $ & $ 0.02 $ & $  0.11 $ & $  9.43 $ & $  0.01 $  \\
 $ 148.93766 $ & $  69.02938 $ & & $ 10.12 $ & $ 0.04 $ & $  0.01 $ & $ 10.00 $ & $ 0.06 $ & $ -0.07 $ & $ 10.04 $ & $ 0.03 $ & $ -0.01 $ & $ 10.16 $ & $ 0.06 $ & $  0.09 $ & $ 10.31 $ & $  0.09 $  \\
 $ 148.67614 $ & $  69.09785 $ & & $ 10.12 $ & $ 0.04 $ & $  0.01 $ & $ 10.02 $ & $ 0.05 $ & $  0.04 $ & $ 10.09 $ & $ 0.04 $ & $  0.01 $ & $ 10.13 $ & $ 0.03 $ & $  0.10 $ & $ 10.05 $ & $  0.12 $  \\
 $ 148.61902 $ & $  69.22280 $ & & $ 10.09 $ & $ 0.04 $ & $  0.02 $ & $ 10.08 $ & $ 0.05 $ & $ -0.01 $ & $ 10.00 $ & $ 0.03 $ & $  0.00 $ & $ 10.10 $ & $ 0.06 $ & $  0.09 $ & $  9.92 $ & $  0.02 $  \\
 $ 148.81385 $ & $  69.25521 $ & & $ 10.14 $ & $ 0.03 $ & $  0.02 $ & $ 10.10 $ & $ 0.03 $ & $ -0.02 $ & $ 10.05 $ & $ 0.03 $ & $ -0.03 $ & $ 10.10 $ & $ 0.04 $ & $  0.02 $ & $ 10.13 $ & $  0.03 $  \\
\dots        & \dots       && \dots  & \dots  & \dots  & \dots  & \dots  & \dots  & \dots  & \dots  & \dots  & \dots  & \dots  & \dots  & \dots  & \dots  \\
\hline  
\hline  
\end{tabular} 
\end{small} 
\end{center}  
\end{table}

\end{landscape}

\clearpage

\begin{landscape}

\begin{table}   
\begin{center}   
\begin{small}  
\caption{Catalog for $16,658$ Point Sources in NGC\,$247$} 
\label{tab:n247}   
\begin{tabular}{rrrrrrrrrrrrrrrrr}   
\\   
\hline  
\hline  
\multicolumn{1}{c}{RA} &
\multicolumn{1}{c}{Dec} &
\multicolumn{1}{c}{} &
\multicolumn{1}{c}{$m_{3.6}$} &
\multicolumn{1}{c}{$\sigma_{3.6}$} &
\multicolumn{1}{c}{$\delta_{3.6}$} &
\multicolumn{1}{c}{$m_{4.5}$} &
\multicolumn{1}{c}{$\sigma_{4.5}$} &
\multicolumn{1}{c}{$\delta_{4.5}$} &
\multicolumn{1}{c}{$m_{5.8}$} &
\multicolumn{1}{c}{$\sigma_{5.8}$} &
\multicolumn{1}{c}{$\delta_{5.8}$} &
\multicolumn{1}{c}{$m_{8.0}$} &
\multicolumn{1}{c}{$\sigma_{8.0}$} &
\multicolumn{1}{c}{$\delta_{8.0}$} &
\multicolumn{1}{c}{$m_{24}$} &
\multicolumn{1}{c}{$\sigma_{24}$} 
\\
\multicolumn{1}{c}{(deg)} &
\multicolumn{1}{c}{(deg)} &
\multicolumn{1}{c}{} &
\multicolumn{1}{c}{(mag)} &
\multicolumn{1}{c}{} &
\multicolumn{1}{c}{} &
\multicolumn{1}{c}{(mag)} &
\multicolumn{1}{c}{} &
\multicolumn{1}{c}{} &
\multicolumn{1}{c}{(mag)} &
\multicolumn{1}{c}{} &
\multicolumn{1}{c}{} &
\multicolumn{1}{c}{(mag)} &
\multicolumn{1}{c}{} &
\multicolumn{1}{c}{} &
\multicolumn{1}{c}{(mag)} &
\multicolumn{1}{c}{} 
\\ 
\hline  
\hline
\dots        & \dots       && \dots  & \dots  & \dots  & \dots  & \dots  & \dots  & \dots  & \dots  & \dots  & \dots  & \dots  & \dots  & \dots  & \dots  \\
 $  11.96215 $ & $ -20.76661 $ & & $  9.92 $ & $ 0.04 $ & $ -0.06 $ & $  9.89 $ & $ 0.04 $ & $  0.01 $ & $  9.98 $ & $ 0.02 $ & $  0.11 $ & $  9.92 $ & $ 0.02 $ & $  0.12 $ & $  9.79 $ & $  0.04 $  \\
 $  11.99306 $ & $ -20.72355 $ & & $ 10.32 $ & $ 0.08 $ & $  0.13 $ & $ 10.10 $ & $ 0.03 $ & $ -0.01 $ & $ 10.14 $ & $ 0.01 $ & $  0.09 $ & $ 10.15 $ & $ 0.05 $ & $  0.11 $ & $ 10.05 $ & $  0.07 $  \\
 $  11.91654 $ & $ -20.89299 $ & & $ 10.39 $ & $ 0.03 $ & $ -0.01 $ & $ 10.43 $ & $ 0.05 $ & $  0.02 $ & $ 10.46 $ & $ 0.02 $ & $  0.11 $ & $ 10.44 $ & $ 0.02 $ & $  0.11 $ & $ 10.43 $ & $  0.03 $  \\
 $  11.73518 $ & $ -20.79068 $ & & $ 10.63 $ & $ 0.06 $ & $ -0.06 $ & $ 10.70 $ & $ 0.04 $ & $ -0.02 $ & $ 10.78 $ & $ 0.02 $ & $  0.09 $ & $ 10.76 $ & $ 0.03 $ & $  0.09 $ & $ 10.70 $ & $  0.08 $  \\
 $  11.85247 $ & $ -20.92059 $ & & $ 10.78 $ & $ 0.06 $ & $ -0.00 $ & $ 10.83 $ & $ 0.05 $ & $  0.01 $ & $ 10.86 $ & $ 0.02 $ & $  0.12 $ & $ 10.82 $ & $ 0.04 $ & $  0.11 $ & $ 10.70 $ & $  0.07 $  \\
\dots        & \dots       && \dots  & \dots  & \dots  & \dots  & \dots  & \dots  & \dots  & \dots  & \dots  & \dots  & \dots  & \dots  & \dots  & \dots  \\
\hline  
\hline  
\end{tabular} 
\end{small} 
\end{center}  
\end{table}

\begin{table}   
\begin{center}   
\begin{small}  
\caption{Catalog for $5,617$ Point Sources in NGC\,$7793$} 
\label{tab:n7793}   
\begin{tabular}{rrrrrrrrrrrrrrrrr}   
\\   
\hline  
\hline  
\multicolumn{1}{c}{RA} &
\multicolumn{1}{c}{Dec} &
\multicolumn{1}{c}{} &
\multicolumn{1}{c}{$m_{3.6}$} &
\multicolumn{1}{c}{$\sigma_{3.6}$} &
\multicolumn{1}{c}{$\delta_{3.6}$} &
\multicolumn{1}{c}{$m_{4.5}$} &
\multicolumn{1}{c}{$\sigma_{4.5}$} &
\multicolumn{1}{c}{$\delta_{4.5}$} &
\multicolumn{1}{c}{$m_{5.8}$} &
\multicolumn{1}{c}{$\sigma_{5.8}$} &
\multicolumn{1}{c}{$\delta_{5.8}$} &
\multicolumn{1}{c}{$m_{8.0}$} &
\multicolumn{1}{c}{$\sigma_{8.0}$} &
\multicolumn{1}{c}{$\delta_{8.0}$} &
\multicolumn{1}{c}{$m_{24}$} &
\multicolumn{1}{c}{$\sigma_{24}$} 
\\
\multicolumn{1}{c}{(deg)} &
\multicolumn{1}{c}{(deg)} &
\multicolumn{1}{c}{} &
\multicolumn{1}{c}{(mag)} &
\multicolumn{1}{c}{} &
\multicolumn{1}{c}{} &
\multicolumn{1}{c}{(mag)} &
\multicolumn{1}{c}{} &
\multicolumn{1}{c}{} &
\multicolumn{1}{c}{(mag)} &
\multicolumn{1}{c}{} &
\multicolumn{1}{c}{} &
\multicolumn{1}{c}{(mag)} &
\multicolumn{1}{c}{} &
\multicolumn{1}{c}{} &
\multicolumn{1}{c}{(mag)} &
\multicolumn{1}{c}{} 
\\ 
\hline  
\hline
\dots        & \dots       && \dots  & \dots  & \dots  & \dots  & \dots  & \dots  & \dots  & \dots  & \dots  & \dots  & \dots  & \dots  & \dots  & \dots  \\
 $ 359.44540 $ & $ -32.47863 $ & & $ 10.50 $ & $ 0.01 $ & $ -0.04 $ & $ 10.73 $ & $ 0.01 $ & $  0.01 $ & $ 10.62 $ & $ 0.01 $ & $ -0.01 $ & $  9.75 $ & $ 0.09 $ & $ -0.90 $ & $ 10.26 $ & $  0.06 $  \\
 $ 359.44747 $ & $ -32.68495 $ & & $ 11.96 $ & $ 0.02 $ & $ -0.02 $ & $ 11.99 $ & $ 0.08 $ & $  0.01 $ & $ 12.98 $ & $ 0.20 $ & $  0.90 $ & $ 11.66 $ & $ 0.02 $ & $ -0.22 $ & $ 11.29 $ & $  0.11 $  \\
 $ 359.35330 $ & $ -32.59285 $ & & $ 12.47 $ & $ 0.03 $ & $ -0.02 $ & $ 12.50 $ & $ 0.01 $ & $  0.03 $ & $ 12.45 $ & $ 0.01 $ & $ 99.99 $ & $ 12.18 $ & $ 0.08 $ & $ -0.35 $ & $ 13.10 $ & $ 99.99 $  \\
 $ 359.52741 $ & $ -32.65571 $ & & $ 12.69 $ & $ 0.09 $ & $ -0.01 $ & $ 12.75 $ & $ 0.03 $ & $  0.08 $ & $ 13.21 $ & $ 0.17 $ & $  0.57 $ & $ 12.29 $ & $ 0.02 $ & $ -0.41 $ & $ 12.62 $ & $ 99.99 $  \\
 $ 359.42730 $ & $ -32.68638 $ & & $ 13.14 $ & $ 0.11 $ & $  0.11 $ & $ 12.92 $ & $ 0.12 $ & $ -0.04 $ & $ 13.13 $ & $ 0.01 $ & $ 99.99 $ & $ 12.70 $ & $ 0.06 $ & $ -0.26 $ & $ 12.54 $ & $ 99.99 $  \\
\dots        & \dots       && \dots  & \dots  & \dots  & \dots  & \dots  & \dots  & \dots  & \dots  & \dots  & \dots  & \dots  & \dots  & \dots  & \dots  \\
\hline  
\hline  
\end{tabular} 
\end{small} 
\end{center}  
\end{table}

\end{landscape}

\clearpage

\end{appendix}


\begin{thebibliography}{27}
\expandafter\ifx\csname natexlab\endcsname\relax\def\natexlab#1{#1}\fi

\bibitem[{{Ashby} {et~al.}(2009)}]{ref:Ashby_2009}
{Ashby}, M.~L.~N. {et~al.} 2009, \apj, 701, 428

\bibitem[{{Barmby} {et~al.}(2006)}]{ref:Barmby_2006}
{Barmby}, P. {et~al.} 2006, ApJ Letters, 650, L45

\bibitem[{{Benjamin} {et~al.}(2003)}]{ref:Benjamin_2003}
{Benjamin}, R.~A. {et~al.} 2003, \pasp, 115, 953

\bibitem[Bolatto et al.(2007)]{ref:Bolatto_2007} Bolatto, A.~D., Simon, 
J.~D., Stanimirovi{\'c}, S., et al.\ 2007, \apj, 655, 212 

\bibitem[{{Bonanos} {et~al.}(2006)}]{ref:Bonanos_2006}
{Bonanos}, A.~Z. {et~al.} 2006, \apj, 652, 313

\bibitem[{{Dale} {et~al.}(2009)}]{ref:Dale_2009}
{Dale}, D.~A. {et~al.} 2009, ApJ, 703, 517

\bibitem[{{Fazio} {et~al.}(2004)}]{ref:Fazio_2004}
{Fazio}, G.~G. {et~al.} 2004, \apjs, 154, 10

\bibitem[{{Gardner} {et~al.}(2006)}]{ref:Gardner_2006}
{Gardner}, J.~P. {et~al.} 2006, \ssr, 123, 485

\bibitem[{{Gerke} \& {Kochanek}(2013)}]{ref:Gerke_2012}
{Gerke}, J.~R. \& {Kochanek}, C.~S. 2013, \apj, 762, 64

\bibitem[{{Gerke} {et~al.}(2011){Gerke}, {Kochanek}, {Prieto}, {Stanek}, \&
  {Macri}}]{ref:Gerke_2011}
{Gerke}, J.~R., {Kochanek}, C.~S., {Prieto}, J.~L., {Stanek}, K.~Z., \&
  {Macri}, L.~M. 2011, \apj, 743, 176

\bibitem[{{Gieren} {et~al.}(2005)}]{ref:Gieren_2005}
{Gieren}, W. {et~al.} 2005, ApJ, 628, 695

\bibitem[{{Gieren} {et~al.}(2006)}]{ref:Gieren_2006}
---. 2006, \apj, 647, 1056

\bibitem[Gordon et al.(2011)]{ref:Gordon_2011} Gordon, K.~D., Meixner, 
M., Meade, M.~R., et al.\ 2011, \aj, 142, 102 

\bibitem[Helou et al.(2004)]{ref:Helou_2004} Helou, G., Roussel, H., 
Appleton, P., et al.\ 2004, \apjs, 154, 253 

\bibitem[{{Kennicutt} {et~al.}(2003)}]{ref:Kennicutt_2003}
{Kennicutt}, Jr., R.~C. {et~al.} 2003, PASP, 115, 928

\bibitem[{{Khan} {et~al.}(2015){Khan}, {Kochanek}, {Stanek}, \&
  {Gerke}}]{ref:Khan_2015}
{Khan}, R., {Kochanek}, C.~S., {Stanek}, K.~Z., \& {Gerke}, J. 2015, \apj

\bibitem[{{Khan} {et~al.}(2013){Khan}, {Stanek}, \& {Kochanek}}]{ref:Khan_2013}
{Khan}, R., {Stanek}, K.~Z., \& {Kochanek}, C.~S. 2013, \apj, 767, 52

\bibitem[{{Khan} {et~al.}(2011){Khan}, {Stanek}, {Kochanek}, \&
  {Bonanos}}]{ref:Khan_2011}
{Khan}, R., {Stanek}, K.~Z., {Kochanek}, C.~S., \& {Bonanos}, A.~Z. 2011, \apj,
  732, 43

\bibitem[{{Khan} {et~al.}(2010){Khan}, {Stanek}, {Prieto}, {Kochanek},
  {Thompson}, \& {Beacom}}]{ref:Khan_2010}
{Khan}, R., {Stanek}, K.~Z., {Prieto}, J.~L., {Kochanek}, C.~S., {Thompson},
  T.~A., \& {Beacom}, J.~F. 2010, ApJ, 715, 1094

\bibitem[{{Madore} {et~al.}(2009){Madore}, {Freedman}, {Catanzarite}, \&
  {Navarrete}}]{ref:Madore_2009}
{Madore}, B.~F., {Freedman}, W.~L., {Catanzarite}, J., \& {Navarrete}, M. 2009,
  \apj, 694, 1237

\bibitem[Martinez-Manso et al.(2015)]{ref:Martinez_2015} Martinez-Manso, 
J., Gonzalez, A.~H., Ashby, M.~L.~N., et al.\ 2015, \mnras, 446, 169 

\bibitem[{{McQuinn} {et~al.}(2007)}]{ref:McQuinn_2007}
{McQuinn}, K.~B.~W. {et~al.} 2007, ApJ, 664, 850

\bibitem[{{Meixner} {et~al.}(2006)}]{ref:Meixner_2006}
{Meixner}, M. {et~al.} 2006, AJ, 132, 2268

\bibitem[Mould et al.(2008)]{ref:Mould_2008} Mould, J., Barmby, P., 
Gordon, K., et al.\ 2008, \apj, 687, 230 

\bibitem[{{Rieke} {et~al.}(2004)}]{ref:Rieke_2004}
{Rieke}, G.~H. {et~al.} 2004, \apjs, 154, 25

\bibitem[{{Saha} {et~al.}(2006)}]{ref:Saha_2006}
{Saha}, A. {et~al.} 2006, \apjs, 165, 108

\bibitem[{{Sheth} {et~al.}(2008)}]{ref:Sheth_2008}
{Sheth}, K. {et~al.} 2008, in Spitzer Proposal ID \#60007, 60007

\bibitem[{{Stetson}(1992)}]{ref:Stetson_1992}
{Stetson}, P.~B. 1992, 25, 297

\bibitem[{{Thompson} {et~al.}(2009){Thompson}, {Prieto}, {Stanek}, {Kistler},
  {Beacom}, \& {Kochanek}}]{ref:Thompson_2009}
{Thompson}, T.~A., {Prieto}, J.~L., {Stanek}, K.~Z., {Kistler}, M.~D.,
  {Beacom}, J.~F., \& {Kochanek}, C.~S. 2009, ApJ, 705, 1364

\bibitem[{{Tully} {et~al.}(2009){Tully}, {Rizzi}, {Shaya}, {Courtois},
  {Makarov}, \& {Jacobs}}]{ref:Tully_2009}
{Tully}, R.~B., {Rizzi}, L., {Shaya}, E.~J., {Courtois}, H.~M., {Makarov},
  D.~I., \& {Jacobs}, B.~A. 2009, \aj, 138, 323

\bibitem[{{Werner} {et~al.}(2004)}]{ref:Werner_2004}
{Werner}, M.~W. {et~al.} 2004, \apjs, 154, 1

\bibitem[Willner et al.(2004)]{ref:Willner_2004} Willner, S.~P., Ashby, 
M.~L.~N., Barmby, P., et al.\ 2004, \apjs, 154, 222 

\end{thebibliography}
\end{document}